\documentstyle{article}
\input amssym.def

\newcommand{\eq}{\begin{equation}}
\newcommand{\eqx}{\end{equation}}
\newcommand{\eqn}{\begin{eqnarray}}
\newcommand{\eqnx}{\end{eqnarray}}
\newcommand{\psiz}{\Psi_{z_0}(z_1,z_2,z_3)}
\newcommand{\dr}{\partial}
\newcommand{\f}[2]{\frac{#1}{#2}}
\newcommand{\Z}{{\Bbb{Z}}}
\newcommand{\C}{{\Bbb{C}}}
\newcommand{\slc}{$SL(2,\C)$}
\newcommand{\slz}{$SL(2,\Z)$}
\newcommand{\lm}{\lambda}
\newcommand{\om}{\omega}
\newcommand{\ra}{\longrightarrow}
\newcommand{\HH}{{\cal{H}}}
\newcommand{\dd}[2]{\f{d^{#1}}{d#2^{#1}}}

\newcommand{\rr}[4]{#1, { #2 \/}{ #3} #4}

\title{Modular Invariance and the Odderon}
\author{Romuald Janik\\
\small Institute of Physics, Jagellonian University, Cracow\\
\small Reymonta 4, 30-059 Krakow\\
\small e-mail: ufrjanik@jetta.if.uj.edu.pl}

\begin{document}

\maketitle

\begin{abstract}
We identify a new symmetry for the equations governing odderon amplitudes,
corresponding in the Regge limit of QCD to the exchange of 3
reggeized gluons. The symmetry is
a  modular invariance with respect to the
unique normal subgroup of \slz {\,} of index 2.
 This leads to a natural
description of the Hamiltonian and conservation-law operators as
acting on the moduli space of elliptic curves
with a fixed ``sign'':  elliptic curves are identified if they can be
transformed into each other by an {\em even} number of Dehn twists.
\end{abstract}


\section{Introduction}

Recent experimental data in  small Bjorken $x$ region
have gained much attention in
 the study of the Regge limit of QCD. Already in the late 70's
the amplitude corresponding to the exchange of
two reggeized gluons was calculated,
and the formula for the intercept of the
renowned BFKL pomeron was derived (see \cite{BFKL} \cite{Lip}).
Lipatov's solution depends in a crucial way
on the global conformal symmetry of the problem. The BFKL equations were
generalized to the case of the exchange of 3 reggeized gluons
 (so-called odderon) by
Kwieci{\'n}ski, Prasza{\l}owicz \cite{KwPrasz} and Bartels \cite{GLLA}.
Later, this approach was extended to the case of arbitrary
number of reggeons
 in the form of the Generalised Leading Logarithm Approximation (GLLA)
\cite{GLLA}.

In the  large $N_c$ limit, the intriguing
connection between the GLLA equations
and exactly solvable lattice models was established. The GLLA equations
are reduced in this limit to a  Schroedinger equation with a two-body
interaction Hamiltonian. Due to the holomorphic separability of this
 Hamiltonian the problem reduces further to
the exactly solvable Heisenberg XXX spin $s=0$ chain
(\cite{KorFad}, \cite{LipSolv}, \cite{Kor}). However, despite the  richness
of mathematical structures involved (global \slc-invariance and abundance of
integral of motions ) as well as diversity of approaches to the problem
(Yang-Baxter equations, Bethe ansatz(\cite{KorFad}, \cite{LipSolv}, \cite{Kor},
\cite{SW}) quasiclassical approximation\cite{QCl}),
 the explicit expression for the intercept of the
odderon has not been found yet.

In this letter we  show that the odderon possesses (albeit in a
somewhat hidden form) the  modular symmetry, ubiquitous in
conformal field theories (CFT)
and string theory. In the next section we recall the theory of
the odderon, then, following Lipatov,
the consequences of global \slc
invariance. As new results, we  analyse the role of cyclic symmetry in
this framework and  we explicitely demonstate the  link to modular
invariance through two alternative descriptions of elliptic curves.
This link may lead to the effective, two-dimensional
string theory for QCD in the moduli space of elliptic curves
with fixed ``parity'', corresponding to the transformations
of the torus through the even number of Dehn twists.

\section{The Odderon}

The Regge limit of QCD is defined as the kinematical region
\eq
s\gg -t \approx M^2
\eqx
where $M$ is the hadron mass scale, or,  in the case of Deep Inelastic
Scattering, as the small $x=Q^2/s$ limit. Here we sketch,
following \cite{Kor}, the equivalence between the Regge intercept of
amplitudes and energy levels of a two-body  Hamiltonian.

The aim is to find the Regge behaviour of the amplitude $A(s,t)\sim
s^{\om_0+1}$. Mellin transformation leads to
\eq
A(s,t)=is\int_{\delta-i\infty}^{\delta+i\infty}\f{d\om}{2\pi i}
\left(\f{s}{M^2}\right)^\om A(\om,t)
\eqx
so the Regge behavior corresponds to finding
the poles of the amplitude $A(\om,t)$.
Rewriting this amplitude as the convolution of ``hadron''  wave
functions $\Phi_{A,B}$ and a kernel $T(\{k_i\},\{k'_j\},\om)$ we get:
\eq
A(\om,t)=\int d^2k_i \int d^2k'_j \Phi_A(\{k_i\})T(\{k_i\},\{k'_j\},\om)
\Phi_B(\{k'_j\})
\eqx
where $\{k_i\}$ and $\{k'_j\}$ are the transverse momenta of the $N$
exchanged reggeons ( in the case of odderon $N=3$ ).
The next step amounts to writing the Bethe-Salpeter equations for the
kernel $T$:
\eq
\om T(\om)=T_0+\HH T(\om)
\eqx
where $T_0$  is the free propagator and $\HH$ is the
operator corresponding to the insertion of single gluonic interactions
between all pairs of reggeons. This equation can be formally solved:
\eq
T(\om)=\f{T_0}{\om-\HH}
\eqx
Therefore the poles of the amplitude correspond to the eigenvalues
of the hamiltonian operator $\HH$. After performing   Fourier transformation
( $k_i\ra b_i$) and  using the complex notation $z_j:=x_j+i
y_j$, the Hamiltonian splits into a sum of a holomorphic part and an
antiholomorphic part. In the large $N_c$ limit the
two commute. It is therefore sufficient  to consider only
the holomorphic part, which
in the case of odderon reads
\eq
(H(z_1,z_2)+H(z_2,z_3)+H(z_3,z_1))\Psi(z_1,z_2,z_3)=E\Psi(z_1,z_2,z_3)
\label{e.hamilt}
\eqx
where
\eq
H(z_1,z_2)=\sum_{l=0}^{\infty}\f{2l+1}{l(l+1)-L^2_{12}}-\f{2}{l+1}
\eqx
with
\eq
L^2_{12}:=-z_{12}^2\f{d}{dz_1}\f{d}{dz_2}
\eqx
being the holomorphic Casimir operator of the group \slc. The eigenvalue $E$
of the holomorphic hamiltonian and the corresponding eigenvalue $\bar{E}$
of the antiholomorphic one are related to the Regge intercept by the formula:
\eq
\om_0=\f{\alpha_s N_c}{4\pi}(E+\bar{E})
\eqx

The celebrated BFKL solution ($N=2$ case) corresponds in this language to
 finding the maximal eigenvalue of the equation
\eq
H(z_1,z_2)\Psi(z_1,z_2)=E\Psi(z_1,z_2)
\eqx
and has the known solution
\eq
E=-4[\psi(m)-\psi(1)]
\label{ccc}
\eqx
where $\psi$ is the derivative of the logarithm of the  Euler $\Gamma$ function
and $m$ is a conformal weight. The maximum of (\ref{ccc}) is achieved at
$m=1/2$ and reproduces the BFKL slope
\eq
\omega_0^{BFKL}=\frac{\alpha_s N_c}{\pi}4\ln 2
\eqx

\subsection{Conservation laws}

\label{s.lipstrat}

The Hamiltonian $H$ is invariant
with respect to the action of \slc {\,}
on holomorphic functions given by:
\eq
(g\cdot \Psi)(z_1,z_2,z_3)=\Psi\left(\f{az_1+b}{cz_1+d},
\f{az_2+b}{cz_2+d},\f{az_3+b}{cz_3+d}\right)
\quad \hbox{ for }g=
\left(\begin{array}{cc}
a&c\\
b&d
\end{array}\right) \in\hbox{\slc}
\eqx
Therefore it commutes with the holomorphic
Casimir operator for this representation:
\eq
\hat{q}_2:=
-z_{12}^2\f{d}{dz_1}\f{d}{dz_2}
-z_{23}^2\f{d}{dz_2}\f{d}{dz_3}
-z_{31}^2\f{d}{dz_3}\f{d}{dz_1}
\eqx
This enables us to consider functions transforming under the unitary
representations of \slc {\,}
labelled by $n\in\Bbb{N}$ and $\nu\in\Bbb{R}$. In this case
 the eigenvalue $q_2$ is $((1+n)/2+i\nu)((-1+n)/2+i\nu)$.

It has been shown \cite{Lipq3} that the system possesses another
integral of motion --- an operator $\hat{q}_3$:
\eq
\hat{q}_3=z_{12}z_{23}z_{31}\dr_1\dr_2\dr_3
\eqx
which commutes with the hamiltonian $H$.

One of the strategies  for solving the odderon problem,
 proposed by Lipatov
\cite{Lipq3}, was to diagonalize the conservation laws $\hat{q}_2$ and
$\hat{q}_3$ and to substitute the solution into the Schroedinger equation
in order to find the energy eigenvalue.

%

\subsection{Conformal ansatz}

Lipatov \cite{LipAns} has chosen an ansatz, which automatically diagonalizes
$\hat{q}_2$:
\eq
\psiz={\left(\f{z_{12}z_{23}z_{31}}{z_{10}^2z_{20}^2z_{30}^2}
\right)}^{m/3}\varphi(\lm)
\eqx
where $m=1/2+i\nu+n/2$, $n$ is an integer and $\nu$ is a real number.
Here, $z_0\in \C$ is just a parameter and
$\lm$ is the anharmonic ratio:
\eq
\lm=\f{z_{12}z_{30}}{z_{13}z_{20}}
\eqx


Lipatov further derived the form of the operator $\hat{q}_3$ within
this ansatz. Inserting $\psiz$ into the equation
\eq
\label{e.cons}
\hat{q}_3\Psi(z_1,z_2,z_3)=q_3\cdot\Psi(z_1,z_2,z_3)
\eqx
and canceling the factor $(\ldots)^{m/3}$ he obtained:
\eq
\label{e.q3lip}
\nabla_1\f{1}{\lm(1-\lm)}\nabla_2\nabla_3\varphi(\lm)=q_3 \varphi(\lm)
\eqx
where
\eqn
\nabla_1 &=& \f{m}{3}(1-2\lm)+\lm(1-\lm)\dr,\\
\nabla_2 &=& \f{m}{3}(1+\lm)+\lm(1-\lm)\dr,\\
\nabla_3 &=& -\f{m}{3}(2-\lm)+\lm(1-\lm)\dr,\\
\eqnx

The Hamiltonian (\ref{e.hamilt}) has also been rewritten in terms of $\lm$.

\section{Cyclic invariance}

It is easy to see that both the Hamiltonian $H$ and $\hat{q}_3$ are
invariant under cyclic permutations of the gluonic coordinates
$z_1,z_2,z_3$. We  show now how this symmetry manifests itself in the formalism
of the preceding section.
Under the permutation $z_1\ra z_2\ra z_3$
the anharmonic ratio transforms as follows:
\eq
\lm \ra 1-\f{1}{\lm} \ra \f{1}{1-\lm}
\eqx
We postulate, that the ground state is symmetric under this
transformation and  so
\eq
\varphi(\lm)=f(s_1,s_2,s_3,\tilde{j})
\eqx
where $s_i$ are the symmetric polynomials in $x_1=\lm, x_2=1-1/\lm$ and
$x_3=1/(1-\lm)$, and $\tilde{j}$ is the Vandermonde determinant.
Namely
\eqn
s_1 &=& x_1+x_2+x_3=\f{\lm^3-3\lm+1}{\lm(\lm-1)} \\
s_2 &=& x_1x_2+x_2x_3+x_3x_1=\f{\lm^3-3\lm^2+1}{\lm(\lm-1)} \\
s_3 &=& x_1x_2x_3= -1\\
\tilde{j} &=& (x_1-x_2)(x_2-x_3)(x_3-x_1) = \f{(\lm^2-\lm+1)^3}{\lm^2
(\lm-1)^2}
\eqnx
It turns out that the only independent quantity is
$A=s_1+s_2=\f{(\lm+1)(2\lm-1)(\lm-2)}{\lm(\lm-1)}$ related to $\tilde{j}$ by
the equation $4\tilde{j}=A^2+27$.
It is convenient to introduce the notation:
\eqn
B&:=&8A=\sqrt{j-1728} \label{e.sqrt}\\
j&:=&256 \tilde{j}=2^8  \f{(\lm^2-\lm+1)^3}{\lm^2(\lm-1)^2}
\label{e.vander}
\eqnx

At this moment we make a refinement of Lipatov's ansatz, namely

\eqn
\psiz&=&{\left(\f{z_{12}z_{23}z_{31}}{z_{10}^2z_{20}^2z_{30}^2}
\right)}^{m/3}f(B)\nonumber \\
&=&
{\left(\f{z_{12}z_{23}z_{31}}{z_{10}^2z_{20}^2z_{30}^2}
\right)}^{m/3}f\left(8\f{(\lm+1)(2\lm-1)(\lm-2)}{\lm(\lm-1)}
\right)
\eqnx
where $m=1/2+i\nu+n/2$, $n$ is an integer and $\nu$ is a real number.
$\lm$ is the anharmonic ratio:
\eq
\lm=\f{z_{12}z_{30}}{z_{13}z_{20}}
\eqx

Now we insert the function $\varphi(\lm)=f(B)$ into the conservation law
(\ref{e.q3lip}). After reexpressing the result in terms of $j$ and
$B=\sqrt{j-1728}$  we get:

\eqn
\label{e.q3ost}
&&\Biggr\{\f{j^2}{2}\dd{3}{B} +2\sqrt{j-1728}j\dd{2}{B}+ (j(1+\f{m(1-m)}{6}
)-3\cdot 2^8)\f{d}{dB}+\nonumber \\
&&\f{(m-3)m^2)}{27}\sqrt{j-1728}-8q_3\Biggl\}f(\sqrt{j-1728})=0
\eqnx

The orginal Hamiltonian (\ref{e.hamilt}) expressed by Lipatov in terms of
$\lm$ can also be recast using the functions $B=\sqrt{j-1728}$ (although
obtaining an explicit expression seems to be highly non-trivial).

In the next section we will show that the variable $j$ can indeed be
considered as a modular invariant and we give a geometric interpretation
of the symmetry considered here.

\section{Elliptic curves}

According to one of the many possible definitions
(see e.g. \cite{elliptic}), an elliptic curve
is a complex curve of genus one. There are two
alternative descriptions of these objects. The first one is the
Weierstrass parametrization which labels each elliptic curve by a
complex number $\lm\in \C$. The curve given by $\lm$ is given by the
equation
\eq
y^2=x(x-1)(x-\lm)
\eqx
where $x$ and $y$ are complex coordinates. Two such
curves are conformally isomorphic if and only if their $j$-invariants
coincide. The $j$-invariant is given by the formula:
\eq
j=2^8\cdot \f{(\lm^2-\lm+1)^3}{\lm^2(\lm-1)^2}
\eqx
Note that this expression is identical to the Vandermonde determinant
considered before (\ref{e.vander}). The only problem that we encounter here
in providing a geometric interpretation to our variables (\ref{e.sqrt}) and
(\ref{e.vander}) is the fact that $\sqrt{j-1728}$ may differ in sign
for isomorphic
elliptic curves. Therefore one must consider elliptic curves with some
additional structure, which we will define after presenting the other
description of genus one curves.

We see here that, since the Hamiltonian can be expressed through the
invariants $\sqrt{j-1728}$, it can be identified in a natural way with
an operator acting on the moduli space of elliptic curves with that
additional structure.

Alternatively one can view elliptic curves as complex tori
$\C/(\Z+\Z\tau)$ parametrized by $\tau\in\C$ in the upper half-plane.
Another way of looking at this quotient space is to consider it as the torus
obtained by identifying opposite edges in the pararellogram bounded by
$0$,$1$,$\tau$ and $1+\tau$.
The $j$-invariant is now a transcendental function of $\tau$. This
description is linked to the preceding one
by the correspondence \cite{gm2}:
\eq
\lm(\tau)=\left(\f{\Theta_2(0;\tau)}{\Theta_3(0;\tau)}\right)^4
\eqx
where $\Theta_2(0;\tau)$ and $\Theta_3(0;\tau)$ are the Jacobi theta
functions.
Moreover,  the symmetry which leaves $j$ invariant corresponds in this
description to modular invariance in the $\tau$-plane i.e.
\eq
j(\tau)=j(\tau')\;\;\Longleftrightarrow
\tau'=\left(\f{a\tau+b}{c\tau+d}\right)
\quad \hbox{ for }
\left(\begin{array}{cc}
a&b\\
c&d
\end{array}\right) \in\hbox{\slz}
\eqx
In our case, the symmetry $\lm \ra 1-\f{1}{\lm} \ra \f{1}{1-\lm}$
corresponds to modular transformations belonging to $\Gamma^2$ ---
the unique normal subgroup of \slz of index 2 \cite{gm2}. This is an
infinite group generated by the matrices
$\left(\begin{array}{cc}
1&2\\
0&1
\end{array}\right)$ and $\left(\begin{array}{cc}
1&-1\\
1&0
\end{array}\right)$.
We still have to define the additional geometric structure on the torus
which is left invariant by the subgroup $\Gamma^2$.

The two elementary Dehn twists, which generate the full modular group,
are associated to the two noncontractible loops of winding number one.
The ``Dehn twist'' operation consists of cutting the torus along the
chosen loop, twisting one boundary by $2\pi$, and gluing it back
(see e.g. \cite{Thiesen}). Although the number of Dehn twists ($n_D$) is
 not well defined, given an isomorphism
corresponding to a modular transformation between equivalent tori,
the parity of number of Dehn twists ($(-1)^{n_D}$) is well defined.
The subgroup $\Gamma^2$ corresponds precisely to the transformations of the
torus through an even number of Dehn twists. We may therefore attach a
kind of ``sign'' to each torus.


Using the correspondence between $\lm$ and $\tau$ one  can reexpress
the Hamiltonian and the integral of motion in terms of $\tau$.

\section{Conclusions}

In this letter we have shown that the odderon problem possesses
 a new symmetry - {\it i.e.}
modular symmetry with respect to $\Gamma^2$ --- an index 2 normal subgroup of
\slz. Expressing  the conservation law (\ref{e.q3lip}) through
modular invariants (\ref{e.q3ost}) leads in a natural way
to the new methods of
solving $3^{rd}$ order Fuchsian differential equations proposed by
B.H. Lian and
S-T Yau's  in the framework of mirror symmetry \cite{Yau}.
 This analogy may be an aid in
carrying out Lipatov's strategy mentioned in section \ref{s.lipstrat} and
may lead to the analytical solution of the odderon problem.\cite{JANIKFUTURE}

Apart from the practical applications of this symmetry, we hope
that it may lead to deeper  understanding of the Regge limit of QCD.
The modular invariance of the odderon leads to a natural
interpretation of all the operators as acting on the moduli space of
genus one curves with fixed `sign', {\it i.e.} the even-parity of the number
of Dehn twists.
In particular this may be a further
step in establishing  the relation between QCD and effective string theory.

\subsection*{Acknowledgments}
I would like to thank Dr. Maciej A. Nowak for
suggesting this investigation and
for fruitful discussions.
I would like to express my gratitude
to Prof. Don Zagier for e-mail correspondence which enhanced my
understanding of the \slz {\,}  structures.

\end{document}